\documentclass[oneside]{article}
\usepackage[T1]{fontenc}
\title{Properties of the Effective Hamiltonian \\
for the System of Neutral Kaons\footnote{ Paper presented at
\textbf{The XXXVI Symposium on Mathematical Physics}, \textit{Poster
session}, Toru\'n, Poland, June 9-12, 2004 . This is shortened
version of \cite{p:44}}}
\author{Justyna Jankiewicz \\
Instytute of Physics\\
University of Zielona Gora \\
Prof. Z. Szafrana 4a, Zielona Gora, Poland\\
J.Jankiewicz@proton.if.uz.zgora.pl }
\date {09-05-2004 - 12-04-2004}
\begin{document}
\bibliographystyle{plain}
\maketitle
\begin{abstract}
We study the properties of time evolution of the $K^{0}-\bar{K}^{0} $ system in
spectral formulation. Within the one--pole model we find the exact form of the
diagonal matrix elements of the effective Hamiltonian for this system. It appears
that, contrary to the Lee--Oehme--Yang (LOY) result, these exact diagonal matrix
elements are different if the total system is CPT--invariant but CP--noninvariant.
\end{abstract}
\section{Introduction}
Following the LOY approach, a nonhermitian Hamiltonian
$H_{\parallel} $ is usually used to study the properties of the
particle-antiparticle unstable system \cite{p:5} - \cite{p:11}
\begin{equation}
 H_{\parallel}\equiv M-\frac{i}{2}\Gamma, \label{j1-5}
\end{equation}
where
\begin{equation}
M=M^{+} \ , \  \Gamma = \Gamma^{+}  \label{j1-6}
\end{equation}
are $(2\times2) $ matrices acting in a two-dimensional subspace
$\mathcal{H}_{\parallel} $ of the total state space $\mathcal{H}$.
The  $M $-matrix is called the mass matrix and  $\Gamma$ is the
decay matrix.  Lee, Oehme and Yang derived their approximate
effective Hamiltonian $H_{\parallel}\equiv H_{LOY} $ by adapting
the one-dimensional Weisskopf-Wigner (WW) method to the
two-dimensional case corresponding to
the neutral kaon system. \\
Almost all properties of this system can be described by solving
the Schr\"{o}dinger-like equation \cite{p:5} - \cite{p:8}
\begin{equation}
i\frac{\partial }{\partial t}|\psi
;t\rangle_{\parallel}=H_{\parallel }|\psi ;t\rangle_{\parallel}, \
\ \  (t\geq t_{0}> -\infty ) \label{j1-7}
\end{equation}
(where we have used  $\hbar =c=1 $) with the initial conditions
\begin{equation}
\parallel |\psi ;t=t_{0}\rangle_{\parallel} \parallel =1,
\ \ |\psi ;t_{0}=0\rangle_{\parallel} =0, \label{j1-8}
\end{equation}
for  $|\psi ;t=t_{0}\rangle_{\parallel} $ belonging to the
subspace of states $\mathcal{H}_{\parallel} $
($\mathcal{H}_{\parallel}\subset \mathcal{H} $), spanned by, e.g.,
orthonormal neutral kaons states  $K^{0} $ and $\bar{K}^{0} $. The
solutions of Eq. (\ref{j1-7}) may be written in a matrix form,
which may be used to define the time evolution operator
$U_{\parallel}(t) $ acting in subspace  $\mathcal{H}_{\parallel} $
\begin{equation}
|\psi ;t\rangle_{\parallel}=U_{\parallel}(t)|\psi
;t_{0}=0\rangle_{\parallel}\equiv U_{\parallel}(t)|\psi
\rangle_{\parallel}, \label{j1-9}
\end{equation}
where
\begin{equation}
|\psi \rangle_{\parallel}\equiv a_{1}|\textbf{1}\rangle +
a_{2}|\textbf{2}\rangle \label{j1-10}
\end{equation}
and  $|\textbf{1}\rangle $ denotes particle "1" -- in the present
case $|K^{0}\rangle $ whereas $|\textbf{2}\rangle $ corresponds to
the antiparticle state for particle ''1'':  $|\bar{K}^{0}\rangle,
$ \ $\langle\textbf{j}|\textbf{k}\rangle =\mathcal{\delta}_{jk}, $
\ $j,k=1,2 $. It is usually assumed that the real parts of the
diagonal matrix elements of $H_{\parallel}$, namely $\Re (\cdot),
$
\begin{equation}
\Re (h_{jj})\equiv M_{jj} \  \  (j=1,2), \label{j1-11}
\end{equation}
where
\begin{equation}
h_{jk}=\langle \textbf{j}|H_{\parallel}|\textbf{k} \rangle \
        \           \     \    (j,k=1,2) \label{j1-12}
\end{equation}
correspond to the masses of the particle "1" and its antiparticle
"2" \cite{p:5} - \cite{p:11}.  $\Im (\cdot) $ is the imaginary
part of $h_{jj}$
\begin{equation}
\Im (h_{jj})\equiv \Gamma_{jj} \ \  \  (j=1,2) \label{j1-13}
\end{equation}
and $\Gamma_{jj}$ are interpreted as the decay widths of the
particles. According to the  standard result of the LOY approach,
in a CPT invariant system, i.e. when
\begin{equation}
\Theta H \Theta^{-1}=H, \label{j1-3}
\end{equation}
(where $\Theta = CPT$, $H=H^{+}$ is the Hamiltonian of the total system under consideration) \\
we have
\begin{equation}
h_{11}^{LOY}=h_{22}^{LOY}. \label{j1-14}
\end{equation}

The universal properties of the unstable particle-antiparticle subsystem described by
the $H$ fulfilling the condition (\ref{j1-3}), may be investigated by using the matrix
elements of the exact $U_{\parallel}, $ instead of the approximate one used in the LOY
theory. The exact $U_{\parallel} $ can be written as follows
\begin{equation}
U_{\parallel}(t)=PU(t)P, \label{j1-17}
\end{equation}
where
\begin{equation}
P\equiv |\textbf{1}\rangle \langle\textbf{1}|+ |\textbf{2}\rangle \langle\textbf{2}|,
\label{j1-18}
\end{equation}
and $U(t) $ is the exact evolution operator acting in the whole state space. This
operator is the solution of the Schr\"{o}dinger equation
\begin{equation}
i\frac{\partial }{\partial t}U(t)|\phi \rangle = HU(t)|\phi
\rangle ,   \  \ U(0)=I. \label{j1-19}
\end{equation}
$I $ is the unit operator in the  $\mathcal{H}$ space and $|\phi \rangle \equiv | \phi
; t_{0}=0\rangle \in \mathcal{H} $ is the initial state of the system. \\
In the
remaining part of the poster we will be using the following matrix representation of
the evolution operator
\begin{equation}
U_{\parallel}(t)\equiv \left( \begin{array}{cc} \textbf{A(t)} &
\textbf{0}
\\ \textbf{0} & \textbf{0}
\end{array} \right), \label{j1-21}
\end{equation}
where  $\textbf{0} $ denotes the zero submatrices of the suitable
dimension, and the  $\textbf{A(t)} $ is a  $(2\times 2) $ matrix
acting in  $\mathcal{H}_{\parallel} $
\begin{equation}
\textbf{A(t)}= \left( \begin{array}[c]{ll} A_{11}(t) & A_{12}(t) \\
 A_{21}(t)& A_{22}(t)
\end{array} \right), \label{j1-22}
\end{equation}
where
\begin{equation} A_{jk}(t)=\langle
\textbf{j}|U_{\parallel}(t)|\textbf{k}\rangle \equiv \langle
\textbf{j}|U(t)|\textbf{k}\rangle \ \ (j,k=1,2). \label{j1-23}
\end{equation}
Assuming that the property (\ref{j1-3}) holds and using the
following definitions
\begin{equation}
\Theta |\textbf{1}\rangle \equiv e^{-i\theta } |\textbf{2}\rangle,
\ \ \ \Theta |\textbf{2}\rangle \equiv e^{-i\theta }
|\textbf{1}\rangle, \label{j1-24}
\end{equation}
it can be shown that
\begin{equation}
A_{11}(t)=A_{22}(t). \label{j1-25}
\end{equation}

A very important relation between the amplitudes $A_{12}(t) $ and
$A_{21}(t) $ follows from the famous Khalfin Theorem \cite{p:16} -
\cite{p:18}
\begin{equation}
r(t)\equiv \frac{A_{12}(t)}{A_{21}(t)}=const\equiv r \  \
\Rightarrow  \  \ |r|=1. \label{j1-26}
\end{equation}
General conclusions concerning the properties of the matrix elements of
$H_{\parallel}$ can be drawn by analyzing the following identity \cite{p:7,p:1}
\begin{equation}
H_{\parallel}(t)\equiv i\frac{\partial \textbf{A}(t)}{\partial
t}[\textbf{A}(t)]^{-1}. \label{j1-37}
\end{equation}

Using Eq. (\ref{j1-37}) we can easily find the general formulae for the diagonal
matrix elements $h_{jj}$,  of $H_{\parallel}(t)$ and next assuming (\ref{j1-3}) and
using relation (\ref{j1-25}) which follows from our earlier assumptions, we get
\begin{equation}
h_{11}(t)-h_{22}(t)=\frac{i}{det\textbf{A}(t)}\Biggl(\frac{\partial
A_{21}(t) }{\partial t}A_{12}(t)-\frac{\partial A_{12}(t)
}{\partial t}A_{21}(t)\Biggl). \label{j1-40}
\end{equation}
In \cite{p:1} it was shown, by using relation (\ref{j1-40}), that this result means
that in the considered case (with CPT conserved) for $t>0$  we get the following
theorem
\begin{equation}
h_{11}(t)-h_{22}(t)=0 \ \ \ \Leftrightarrow \ \ \
\frac{A_{12}(t)}{A_{21}(t)}=const \ \ \ (t>0) . \label{j1-42}
\end{equation}
Thus, for  $t>0 $ the problem under study is reduced to the
Khalfin Theorem (see relation (\ref{j1-26})) \cite{p:1}.

Having noticed this,  let us now turn our attention to the conclusions following from
Khalfin's Theorem. $CP $ noninvariance requires that  $|r|\neq 1$
\cite{p:5,p:6,p:7,p:8,p:16,p:18,p:14,p:2}. This means that in this case the following
condition must be fulfilled: $r=r(t)\neq const. $ Consequently, if in the considered
system property (\ref{j1-3}) holds, but at the same time
\begin{equation}
[\mathcal{CP},H]\neq 0  \label{j1-45}
\end{equation}
and the unstable states "1" i "2"  are connected by (\ref{j1-24}),
then in this system for $t>0 $  \cite{p:1}
\begin{equation}
h_{11}(t)-h_{22}(t)\neq 0. \label{j1-46}
\end{equation}
So, in the exact quantum theory the difference $(h_{11}(t)-h_{22}(t))$  cannot be
equal to zero with CPT conserved and CP violated.
\section{A model: one pole approximation}
While describing the two and three pion decay we are mostly interested in the
$|K_{S}\rangle $ and $|K_{L}\rangle $ superposition of $|K^{0}\rangle$ and
$|\bar{K}^{0}\rangle.$ These states correspond to the physical $|K_{S}\rangle $ and
$|K_{L}\rangle $ neutral kaon states \cite{p:2,p:26}
\begin{eqnarray}
|K_{S}\rangle =p|K^{0}\rangle +q|\bar{K}^{0}\rangle,
\ \ |K_{L}\rangle =p|K^{0}\rangle
-q|\bar{K}^{0}\rangle. \label{j1-47}
\end{eqnarray}
Using the spectral formalism we can write an unstable state $|\lambda (t)\rangle $ as
\begin{eqnarray}
|\lambda (t)\rangle =\sum_{q}|q (t)\rangle \omega_{\lambda}(q),
\label{j1-62}
\end{eqnarray}
where $|q (t)\rangle =e^{-itH}|q \rangle $, vectors $|q \rangle $
form a complete set of eigenvectors of the hermitian,
quantum-mechanical Hamiltonian $H$ and $ \omega_{\lambda}(q) =
\langle q|\lambda \rangle.$ If the continuous eigenvalue is
denoted by $m$, we can define the survival amplitude  $A(t)$ (or
the transition amplitude in the case of $K^{0}\leftrightarrow
\bar{K}^{0} $ ) in the following way:
\begin{eqnarray}
A(t)=\int\limits_{Spec(H)}dm \; e^{-imt}\rho (m), \label{j1-62a}
\end{eqnarray}
where the integral extends over the whole spectrum of the
Hamiltonian and density $\rho (m)$ is defined as follows
\begin{eqnarray}
\rho (m)=|\omega_{\lambda}(m)|^{2}, \label{j1-62b}
\end{eqnarray}
where $\omega_{\lambda}(m)=\langle m |\lambda \rangle . $

In accordance with formula (\ref{j1-62}) the unstable
 states $K_{S} $ and $K_{L} $ may
now be written as a superposition of the eigenkets
\begin{eqnarray}
|K_{S}\rangle =\int_{0}^{\infty }dm \;
\sum_{\alpha}\omega_{S,\alpha}(m)|\phi_{\alpha}(m)\rangle ;\label{j1-65}\\
|K_{L}\rangle =\int_{0}^{\infty }dm \;
\sum_{\beta}\omega_{L,\beta}(m)|\phi_{\beta}(m)\rangle
.\label{j1-66}
\end{eqnarray}

The Breit-Wigner ansatz \cite {p:27}
\begin{eqnarray}
\rho_{WB}(m)=\frac{\Gamma}{2\pi
}\frac{1}{(m-m_{0})^{2}+\frac{\Gamma^{2}}{4}} \equiv |\omega
(m)|^{2} \label{j1-63}
\end{eqnarray}
leads to the well known exponential decay law which follows from
the survival amplitude
\begin{eqnarray}
A_{BW}(t)=\int_{-\infty}^{\infty}dm \;
e^{-imt}\rho_{WB}(m)=e^{-im_{0}t}e^{-\frac{1}{2}\Gamma|t|}.
\label{j1-64}
\end{eqnarray}
(Note that the existence of the ground state induces
non-exponential corrections to the decay law and to the survival
amplitude (\ref{j1-64}) --- see \cite{p:2} ). It is therefore
reasonable to assume a suitable form for $\omega_{S,\beta } $ and
$\omega_{L,\beta }$. More specifically, we write \cite{p:2}
\begin{equation}
\omega_{S,\beta}(m)= \sqrt{\frac{\Gamma_{S}}{2\pi}}
\frac{A_{S,\beta}(K_{S}\rightarrow\beta)}{m-m_{S}+ i\frac{\Gamma_{S}}{2}}, \ \
\omega_{L,\beta }(m)=\sqrt{\frac{\Gamma_{L}}{2\pi}}
\frac{A_{L,\beta}(K_{L}\rightarrow\beta)}{m-m_{L}+ i\frac{\Gamma_{L}}{2}}
\label{j1-73}
\end{equation}
where  $A_{S,\beta} $ and  $A_{L,\beta} $ are decay (transition) amplitudes, end thus
\begin{equation}
\rho_{x,\beta}(m)= \frac{\Gamma_{x}}{2\pi}
\frac{(A_{x,\beta}(K_{x}\rightarrow\beta))^{2}}{(m-m_{x})^{2}+
\frac{(\Gamma_{x})^{2}}{4}}, \label{j1-73a}
\end{equation}\
where $x=L,S$.

In the one-pole approximation (\ref{j1-73}) $A_{K^{0}K^{0}}(t) $ can be conveniently
written as
\begin{eqnarray}
A_{K^{0}K^{0}}(t)\nonumber&=&A_{\bar{K}^{0}\bar{K}^{0}}(t)=\nonumber
\\
&=&-\frac{1}{2\pi}\Biggl\{e^{-im_{S}t}
\left(-\int_{0}^{-\frac{m_{S}}{\gamma_{S}}}dy
\frac{e^{-i\gamma_{S}ty}}{y^{2}+1} +\int_{0}^{\infty}dy
\frac{e^{-i\gamma_{S}ty}}{y^{2}+1}\right)+\nonumber\\
&&+e^{-im_{L}t} \left(-\int_{0}^{-\frac{m_{L}}{\gamma_{L}}}dy
\frac{e^{-i\gamma_{L}ty}}{y^{2}+1}+ \int_{0}^{\infty}dy
\frac{e^{-i\gamma_{L}ty}}{y^{2}+1}\right)\Biggl\}. \label{j1-77}
\end{eqnarray}
Collecting  only exponential terms in (\ref{j1-77}) one obtains an
expression analogous to the WW approximation
\begin{equation}
A_{K^{0}K^{0}}(t)=A_{\bar{K}^{0}\bar{K}^{0}}(t)=
\frac{1}{2}\left(e^{-im_{S}t}e^{-\gamma_{S}t}+
e^{-im_{L}t}e^{-\gamma_{L}t}\right)+N_{K^{0}K^{0}}(t).
\label{j1-78}
\end{equation}
Here $N_{K^{0}K^{0}}(t)$ denotes all non-oscillatory terms present in the integral
(\ref{j1-77}).
\section{Diagonal matrix elements of the effective\\ Hamiltonian}
Using the decomposition of type (\ref{j1-78}) and the one-pole
ansatz (\ref{j1-73}), we find the difference (\ref{j1-46}), which
is now  formulated for the $K^{0} - \bar{K}^{0} $ system. Here it
has the following form:
\begin{eqnarray}
h_{11}(t)-h_{22}(t)=\frac{X(t)}{Y(t)}, \label{j1-79}
\end{eqnarray}
where
\begin{eqnarray}
X(t)= i\left(\frac{\partial A_{\bar{K}^{0}K^{0}}(t)}{\partial
t}A_{K^{0}\bar{K}^{0}}(t)-\frac{\partial
A_{K^{0}\bar{K}^{0}}(t)}{\partial t}A_{\bar{K}^{0}K^{0}}(t)\right)
\label{j1-79a}
\end{eqnarray}
and
\begin{eqnarray}
Y(t)= A_{K^{0}K^{0}}(t)A_{\bar{K}^{0}\bar{K}^{0}}(t)
-A_{K^{0}\bar{K}^{0}}(t)A_{\bar{K}^{0}K^{0}}(t). \label{j1-79b}
\end{eqnarray}

Using the above mentioned spectral formulae in the one - pole approximation
(\ref{j1-73})  we get $A_{K^{0}\bar{K}^{0}}(t) $ and
 $A_{\bar{K}^{0}K^{0}}(t) $
\begin{eqnarray}
\nonumber&&A_{K^{0}\bar{K}^{0}}(t)= \frac{1+\pi }{8\pi
p^{\ast}q}\Biggl\{e^{-im_{S}t}e^{-\gamma_{S}t} \Biggl[
1+\nonumber\\
&&+\frac{\sqrt{\gamma_{S}\gamma_{L}}}{ \gamma_{S}} \Biggl(-2 \, i
\, \gamma_{S}C_{I} +D'_{I}-F'_{I}\Biggl)
\Biggl]+\nonumber\\
&&+e^{-im_{L}t}e^{-\gamma_{L}t}
\Biggl[-1+\nonumber\\
&&+\frac{\sqrt{\gamma_{S}\gamma_{L}}}{ \gamma_{L}} \Biggl(2 \, i
\, \gamma_{L}C_{I}-D'_{I}+F'_{I}\Biggl)
\Biggl]\Biggl\}+\nonumber\\
&&+N_{K^{0}\bar{K}^{0}}(t) \label{j1-87}
\end{eqnarray}
and
\begin{eqnarray}
\nonumber&&A_{\bar{K}^{0}K^{0}}(t)= \frac{1+\pi }{8\pi
pq^{\ast}}\Biggl\{e^{-im_{S}t}e^{-\gamma_{S}t} \Biggl[
1+\nonumber\\
&&+\frac{\sqrt{\gamma_{S}\gamma_{L}}}{ \gamma_{S}} \Biggl(2 \, i
\, \gamma_{S}C_{I}-D'_{I}+F'_{I}\Biggl)
\Biggl]+\nonumber\\
&&+e^{-im_{L}t}e^{-\gamma_{L}t}
\Biggl[-1+\nonumber\\
&&+\frac{\sqrt{\gamma_{S}\gamma_{L}}}{ \gamma_{L}} \Biggl(-2 \, i
\, \gamma_{L}C_{I}+D'_{I}-F'_{I}\Biggl)
\Biggl]\Biggl\}+\nonumber\\
&&+N_{\bar{K}^{0}K^{0}}(t), \label{j1-89}
\end{eqnarray}
where $N_{K^{0}\bar{K}^{0}}(t)$, $N_{\bar{K}^{0}K^{0}}(t)$ denotes all non-oscillatory
terms and $C_{I}, D'_{I}, F'_{I}$ are defined in \cite{p:2}.

Using the expression  for the derivative of $E_{i}$ we can find the derivatives which
will be necessary for the following calculations $\frac{\partial
A_{K^{0}\bar{K}^{0}}(t)}{\partial t} $ and $\frac{\partial
A_{\bar{K}^{0}K^{0}}(t)}{\partial t} :$
\begin{eqnarray}
\frac{\partial A_{K^{0}\bar{K}^{0}}(t)}{\partial
t}\nonumber&=&\frac{1+\pi } {8\pi
p^{\ast}q}\Biggl\{e^{-im_{S}t}e^{-\gamma_{S}t}
\Biggl[-im_{S}-\gamma_{S}+\nonumber\\
&&+\sqrt{\gamma_{S}\gamma_{L}} \Biggl(2 i \gamma_{S}
C_{I}-D'_{I}+F'_{I}\Biggl)\Biggl]+\nonumber\\
&&+e^{-im_{L}t} e^{-\gamma_{L}t}\Biggl[im_{L}-\gamma_{L}+\nonumber\\
&&+\sqrt{\gamma_{S}\gamma_{L}} \Biggl(-2 i \gamma_{L}
C_{I}+D'_{I}-F'_{I}\Biggl)\Biggl]\Biggl\}+\nonumber\\
&&+\Delta N_{K^{0}\bar{K}^{0}}(t) \label{j1-91}
\end{eqnarray}
and
\begin{eqnarray}
\frac{\partial A_{\bar{K}^{0}K^{0}}(t)}{\partial
t}\nonumber&=&\frac{1+\pi } {8\pi
pq^{\ast}}\Biggl\{e^{-im_{S}t}e^{-\gamma_{S}t}
\Biggl[-im_{S}-\gamma_{S}+\nonumber\\
&&+\sqrt{\gamma_{S}\gamma_{L}} \Biggl(-2 i \gamma_{S}
C_{I}+D'_{I}-F'_{I}\Biggl)\Biggl]+\nonumber\\
&&+e^{-im_{L}t} e^{-\gamma_{L}t}\Biggl[im_{L}-\gamma_{L}+\nonumber\\
&&+\sqrt{\gamma_{S}\gamma_{L}} \Biggl(2 i \gamma_{L}
C_{I}-D'_{I}+F'_{I}\Biggl)\Biggl]\Biggl\}+\nonumber\\
&&+\Delta N_{\bar{K}^{0}K^{0}}(t), \label{j1-93}
\end{eqnarray}
where $\Delta N_{K^{0}\bar{K}^{0}}(t)$, $\Delta
N_{\bar{K}^{0}K^{0}}(t)$ denotes all non-oscillatory terms.

The states $|K_{L}\rangle$ and $|K_{S}\rangle$ are superpositions of $|K^{0}\rangle$
and $|\bar{K}^{0}\rangle$. The lifetimes of particles $|K_{L}\rangle$ and
$|K_{S}\rangle$ may be denoted by  $\tau_{L}$ and $\tau_{S}$, respectively,
$\tau_{L}=\frac{1}{\gamma_{L}}=5,183\cdot 10^{-8}s$ being much longer than
$\tau_{S}=\frac{1}{\gamma_{S}}=0,8923\cdot 10^{-10}s.$

Below we calculate the difference  (\ref{j1-79}) for $t\sim
\tau_{L}$
\begin{eqnarray}
h_{11}(t\sim \tau_{L})-h_{22}(t\sim
\tau_{L})=\frac{X(t\sim\tau_{L})}{Y(t\sim\tau_{L})}. \label{j1-95}
\end{eqnarray}
If we only consider the long living states $|K_{L}\rangle$ we may drop all the terms
containing $e^{-\gamma_{S}t}|_{t\sim \tau_{L}}$ as they are negligible in comparison
with elements involving the factor $e^{-\gamma_{L}t}|_{t\sim \tau_{L}}.$   We also
drop all the non-oscillatory terms $N_{K^{0}K^{0}}(t),$ $N_{\bar{K}^{0}K^{0}}(t)$,
$N_{K^{0}\bar{K}^{0}}(t)$ present in $A_{K^{0}K^{0}}(t)$, $A_{\bar{K}^{0}K^{0}}(t)$
and $A_{K^{0}\bar{K}^{0}}(t),$ that is in integrals (\ref{j1-77}), (\ref{j1-87}) and
(\ref{j1-89}), because they are extremally small in the region of time  $ t\sim
\tau_{L}$ \cite{p:2,p:28,p:29}. Similarly, because of the properties of the
exponential integral function $E_{i},$ we can drop terms like $\Delta
N_{\bar{K}^{0}K^{0}}$ and $\Delta N_{K^{0}\bar{K}^{0}}$  present in $\frac{\partial
A_{\bar{K}^{0}K^{0}}}{\partial t}$ (\ref{j1-91}) and $\frac{\partial
A_{K^{0}\bar{K}^{0}}}{\partial t}$ (\ref{j1-93}).This conclusion follows from the
asymptotic properties of the exponential integral function $E_{i}$ and the fact that
$\Delta N_{\bar{K}^{0}K^{0}},$ $\Delta N_{K^{0}\bar{K}^{0}}$ only contain expressions
proportional to  $E_{i}$.

We may now calculate the products
$A_{K^{0}K^{0}}(t)A_{\bar{K}^{0}\bar{K}^{0}}(t), $
$A_{K^{0}\bar{K}^{0}}(t)A_{\bar{K}^{0}K^{0}}(t), $\\
$\frac{\partial A_{\bar{K}^{0}K^{0}}}{\partial
t}(t)A_{K^{0}\bar{K}^{0}}(t), $  $\frac{\partial
A_{K^{0}\bar{K}^{0}}}{\partial t}(t)A_{\bar{K}^{0}K^{0}}(t)$,
which, after using the above mentioned properties of
$N_{K^{0}K^{0}}(t)$, $\Delta N_{K^{0}K^{0}}(t)$ and performing
some algebraic transformations, leads to the following form of the
difference (\ref{j1-95}):
\begin{eqnarray}
h_{11}(t\sim \tau_{L})-h_{22}(t\sim\tau_{L}))=
\Biggl(\frac{2\pi^{2}\sqrt{\gamma_{S}\gamma_{L}}}{
\pi^{2}+2\pi+1}\Biggl)\cdot \frac{Z}{W}\neq 0, \label{j1-97a}
\end{eqnarray}
where
\begin{eqnarray}
Z=\nonumber&& 4|p|^{2}|q|^{2}-\frac{\pi^{2}+2\pi
+1 }{4\pi^{2}}\Biggl[1+\nonumber\\
&&+\gamma_{S}\Biggl(4\gamma_{L}C_{I}^{2}+
\frac{1}{\gamma_{L}}(-D_{I}^{'2} -F_{I}^{'2}
+4D_{I}^{'}F_{I}^{'})+\nonumber\\
&&+4 i C_{I}(D_{I}^{'}-F_{I}^{'})\Biggl)\Biggl]\neq
0\label{j1-97b}
\end{eqnarray}
\begin{eqnarray}
W=\nonumber&&2\Biggl(-C_{I}m_{L}+D_{I}^{'}-
F_{I}^{'}\Biggl)+\nonumber\\
&&+i\Biggl[-4C_{I}\gamma_{L}+\frac{m_{L}}{\gamma_{L}}
\Biggl(-D_{I}^{'}+F_{I}^{'}\Biggl)\Biggl]\neq 0. \label{j1-97c}
\end{eqnarray}
\section{Final remarks}
\begin{itemize}
  \item Our results presented in the present poster have shown that in a CPT invariant and CP
noninvariant system in the case of the exactly solvable one-pole model, the diagonal
matrix elements do not have to be equal. In the general case the diagonal elements
depend on time and their difference, for example at $t\sim \tau_{L}$, is different
from zero.  Z and W in (\ref{j1-97a}) are different from zero, so the difference
$(h_{11}(t)-h_{22}(t))|_{t\sim\tau_{L}}\neq 0.$ From this observation a conclusion of
major importance can be drawn, namely that the measurement of the mass difference
$(m_{K^{0}}-m_{\bar{K}^{0}}) $ should not be used while designing CPT invariance
tests. This runs counter to the general conclusions following from the Lee, Oehme and
Yang theory.
  \item A detailed analysis of $h_{jk}(t)$, $(j,k = 1,2)$ shows that the
non-oscillatory elements $N_{\alpha ,\beta}(t), \Delta N_{\alpha ,\beta}(t)$ (where
$\alpha , \beta = K^{0}, {\overline{K}}^{0}$) is the source of the non-zero difference
$(h_{11}(t) - h_{22}(t))$ in the model considered. It is not difficult to verify that
dropping all the terms of  $N_{\alpha ,\beta}(t), \Delta N_{\alpha ,\beta}(t)$ type in
the formula for $(h_{11}(t) - h_{22}(t))$ gives $(h_{11}^{osc}(t) - h_{22}^{osc}(t)) =
0$, where $h_{jj}^{osc}(t)$, $(j = 1,2)$, stands for $h_{jj}(t)$ without the
non-oscillatory terms.
  \item The result  ($h_{11}(t)-h_{22}(t))\neq 0$ seems to be very
important as it has been obtained within the exactly solvable one-pole model based on
the Breit-Wigner ansatz, i.e. the same
model as used by Lee, Oehme and Yang. \\
\end{itemize}
\section*{Acknowledgements}
The author wishes to thank Professor Krzysztof Urbanowski for many
helpful discussions.

\end{document}